\documentclass[aps,prb,amsmath,amssymb,reprint,superscriptaddress]{revtex4-2}

\usepackage{graphicx}
\usepackage{amsmath,amssymb}
\usepackage{xcolor}
\usepackage{color,soul}

\begin{document}

\title{Structure Prediction and Bonding Analysis of B$_{18}$Ag$_2$ Clusters Featuring Double-Ring Motifs}

\author{P. L. Rodr\'iguez-Kessler}
\email{plkessler@cio.mx}
\affiliation{Centro de Investigaciones en \'Optica A.C., Loma del Bosque 115, Lomas del Campestre, Leon, 37150, Guanajuato, Mexico}

\date{\today}

\begin{abstract}
The structural stability, electronic structure, and bonding characteristics of the silver-doped boron cluster B$_{18}$Ag$_2$ were investigated using density functional theory (DFT) combined with global optimization techniques. Basin-hopping searches identify a bent double-ring structure as the global minimum, consisting of two stacked B$_9$ rings symmetrically stabilized by Ag atoms located above and below the boron framework. The UV–Vis absorption spectrum exhibits weak transitions in the near-infrared region and intense bands in the visible and near-ultraviolet regions, reflecting delocalized electronic excitations within the boron framework. Charge analysis indicates moderate electron redistribution from Ag atoms to the boron scaffold. Real-space bonding analyses based on the electron localization function (ELF), reduced density gradient (RDG), and molecular electrostatic potential (MEP) reveal that bonding is dominated by $\sigma$-delocalization over the boron skeleton, while Ag–B interactions are weak, non-directional, and primarily electrostatic. The continuous annular electron delocalization within the double-ring structure suggests an aromatic-like character. These findings establish B$_{18}$Ag$_2$ as a silver-stabilized boron double-ring cluster in which global electron delocalization governs structural stability, while Ag atoms act as axial stabilizing centers that modulate the electronic structure. This work provides new insight into the role of coinage-metal doping in stabilizing extended boron nanostructures.
\end{abstract}

\maketitle

\section{Introduction}

Boron clusters have attracted sustained interest in cluster chemistry owing to their remarkable structural diversity and the prevalence of electron-deficient bonding that gives rise to unique bonding patterns.\cite{doi:10.1126/science.196.4294.1047} Medium-sized boron clusters, in particular, often adopt planar, quasi-planar, or tubular configurations, with the B$_{18}$ system representing an important size regime where extended delocalization and structural flexibility coexist.\cite{doi:10.1021/jp8087918,C6SC02623K} Such frameworks provide an ideal platform for metal doping, which can further stabilize boron skeletons while introducing new electronic and bonding features.
\\
The incorporation of transition and coinage metals into boron clusters has led to a wide range of novel structural motifs, including inverse-sandwich configurations, metal-stabilized boron wheels, and three-dimensional cage-like assemblies.\cite{JIA2014128,Zhuan-Yu2014,PHAM2019186,C5CP01650A,LI202325821,RODRIGUEZKESSLER2025117486} In these systems, metal atoms often interact with the boron framework through multicenter bonding and charge redistribution, promoting enhanced stability and electronic delocalization.\cite{doi:10.1021/acs.inorgchem.7b02585,doi:10.1021/acs.jpclett.0c02656} Despite these advances, doubly doped boron clusters of the form B$_{18}$M$_2$ remain relatively unexplored, particularly for coinage metals, where relativistic effects, d–s hybridization, and flexible coordination environments may give rise to unconventional bonding patterns.\\
Silver doping is especially appealing due to its moderate electronegativity, accessible d orbitals, and ability to participate in both covalent and delocalized bonding interactions. These characteristics suggest that Ag atoms may effectively stabilize extended boron frameworks while preserving or enhancing multicenter electron delocalization. Furthermore, coinage-metal doping has been shown to induce unique electronic properties, including global aromaticity, charge-separated bonding, and metal-mediated delocalization in boron-based nanostructures.\cite{D5CP01078K,GUEVARAVELA2025115487,RODRIGUEZKESSLER2023116538,RODRIGUEZKESSLER2024122062}
\\
In this context, Ag-doped boron clusters have attracted increasing attention due to their relatively weak and flexible metal–boron interactions, as exemplified by systems such as AgB$_8^-$.\cite{D5SC08598E} However, despite these advances, the interaction of Ag atoms with larger boron clusters remains largely unexplored, particularly for systems with multiple metal dopants where cooperative effects may lead to new structural motifs and bonding patterns. Understanding the interplay between metal–boron coordination, charge transfer, and multicenter bonding in such systems is crucial for rationalizing their stability and guiding the design of new metal-doped boron nanostructures with tailored properties.
\\
In this work, we present a comprehensive computational investigation of the B$_{18}$Ag$_2$ cluster using global structural searches combined with density functional theory calculations. The low-energy structures, vibrational stability, and electronic properties are analyzed through charge distribution, electron density descriptors, and bonding analyses. Our results indicate that B$_{18}$Ag$_2$ forms a stable double-ring boron framework capable of accommodating two silver atoms through symmetric multicenter interactions, highlighting the role of coinage-metal doping in stabilizing extended boron nanostructures. These findings broaden the understanding of metal-doped boron clusters and provide further insights into the design of novel boron-based nanomaterials.\cite{C7CP04158F,OLALDELOPEZ2024}

\section{Computational Details}

Calculations performed in this work are carried out by using density functional theory (DFT) as implemented in the ORCA 6.0.0 code.\cite{10.1063/5.0004608} The exchange and correlation energies are addressed by the PBE0 functional in conjunction with the Def2-TZVP basis set.\cite{10.1063/1.478522,B508541A} The RIJCOSX approximation was employed to accelerate the evaluation of the Coulomb and exchange integrals.\cite{doi:10.1021/ct100199k} The default numerical integration grid in ORCA was employed, corresponding to Grid4 for the SCF iterations and a finer grid (Final Grid5) for the final energy evaluation. Atomic positions are self-consistently relaxed through a Quasi-Newton method employing the BFGS algorithm. The SCF convergence criterion is set to TightSCF in the input file. This results in geometry optimization settings of 1.0e$^{-08}$ Eh for total energy change. The  Van  der  Waals  interactions  are  included in the exchange-correlation functionals with empirical dispersion corrections of Grimme DFT-D3(BJ). The electron localization function (ELF) was computed and analyzed using Multiwfn.\cite{https://doi.org/10.1002/jcc.22885} Global minima searches were conducted using standard basin-hopping (BH) algorithm with random rotational–translational perturbations and subsequent DFT local optimization at the PBE0/def2-SVP level.\cite{Rodriguez-Kessler2026,Ortega-Flores2025} Low-lying candidates ($<$20 kcal/mol) were reoptimized at PBE0/def2-TZVP level. Vibrational frequency calculations confirmed all minima as true stationary points (no imaginary modes). Spin states from singlet to quintet were evaluated. Spatial region analyses (calc. grid data) were calculated with Multiwfn.

\section{Results and Discussion}

\begin{figure}[ht]
\centering
\includegraphics[width=0.45\textwidth]{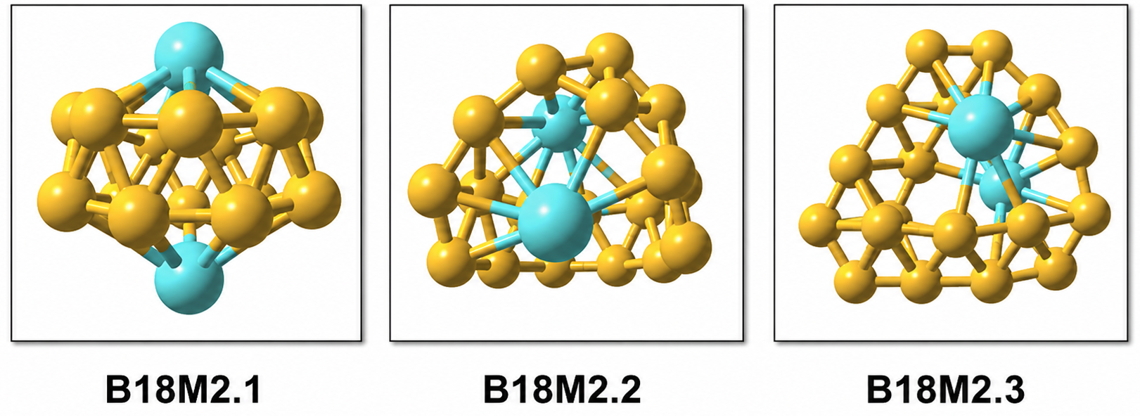}
\caption{Lowest energy structures for B$_{18}$Ag$_2$ cluster at PBE0/def2-TZVP.}
\label{fig_geom}
\end{figure}

The low-energy structures of the B$_{18}$Ag$_2$ cluster reveal a diverse structural landscape in which both layered and cage-like motifs compete for stability (see Fig.~\ref{fig_geom}). The global minimum ({\bf 18M2.1}) adopts a bent double-ring configuration, where two quasi-planar B$_9$ rings are stacked and stabilized by two Ag atoms aligned along the principal axis. In this structure, the silver atoms are located above and below the boron framework, interacting symmetrically with both rings and maximizing coordination while preserving a high degree of structural order.

The higher-energy isomers ({\bf 18M2.2} and {\bf 18M2.3}) exhibit significant deviations from this ideal double-ring arrangement. In particular, {\bf 18M2.2} can be described as a distorted double-ring isomer, in which the boron layers are partially tilted and the Ag atoms are displaced from the central axis, leading to reduced symmetry and weaker coordination. This distortion modifies the inter-ring coupling and results in a moderate increase in energy relative to the global minimum.

In contrast, {\bf 18M2.3} adopts a cage-like structure, where the boron atoms reorganize into a more compact three-dimensional framework. In this configuration, the Ag atoms are more embedded within the boron network, interacting with a larger number of neighboring atoms. This geometry enhances local coordination but sacrifices the clear layered character observed in the global minimum.

Although numerous intermediate structures were identified during the global search, the representative isomers shown here capture the essential structural diversity of the B$_{18}$Ag$_2$ cluster, ranging from layered double-ring motifs to fully three-dimensional cage-like geometries. Overall, the structural evolution reflects a balance between delocalized multicenter B–B bonding within the boron framework and Ag–B interactions that modulate geometry and stability, giving rise to several energetically competitive isomers.

\subsection{Relative energies}

The relative energies clearly indicate a strong preference for the double-ring structure, which is identified as the global minimum at both the PBE0 and $\omega$B97X-D3 levels of theory, underscoring its superior structural stability. In contrast, the distorted double-ring and cage-like isomers are significantly less stable, lying higher in energy by approximately 0.93–2.05~eV, which suggests that they are thermodynamically inaccessible under equilibrium conditions. Notably, the excellent agreement between PBE0 and $\omega$B97X-D3 confirms that the isomer energy ordering is insensitive to the choice of exchange–correlation functional, thereby reinforcing the double-ring motif as the dominant structural feature for this cluster composition.

\begin{table}[h]
\centering
\setlength{\tabcolsep}{12pt} 
\caption{Relative energies of the lowest B$_{18}$Ag$_2$ isomers (in eV). The global minimum is denoted by GM.}
\begin{tabular}{lcc}
\hline\hline
Isomer & PBE0 & $\omega$B97X-D3 \\
\hline
Bent double-ring (GM) & 0.0 & 0.0 \\
Distorted double-ring & 0.93 & 0.54 \\
Cage-like isomer & 2.05 & 1.91 \\
\hline\hline
\end{tabular}
\end{table}

\subsection{Vibrational and Optical Properties}

\begin{figure*}[ht]
\centering
\begin{tabular}{cc}
 \resizebox*{0.40\textwidth}{!}{\includegraphics{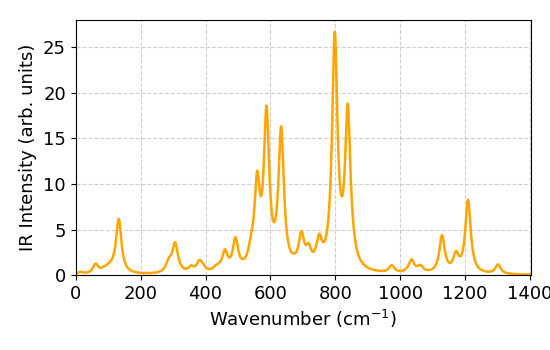}} &
\resizebox*{0.39\textwidth}{!}{\includegraphics{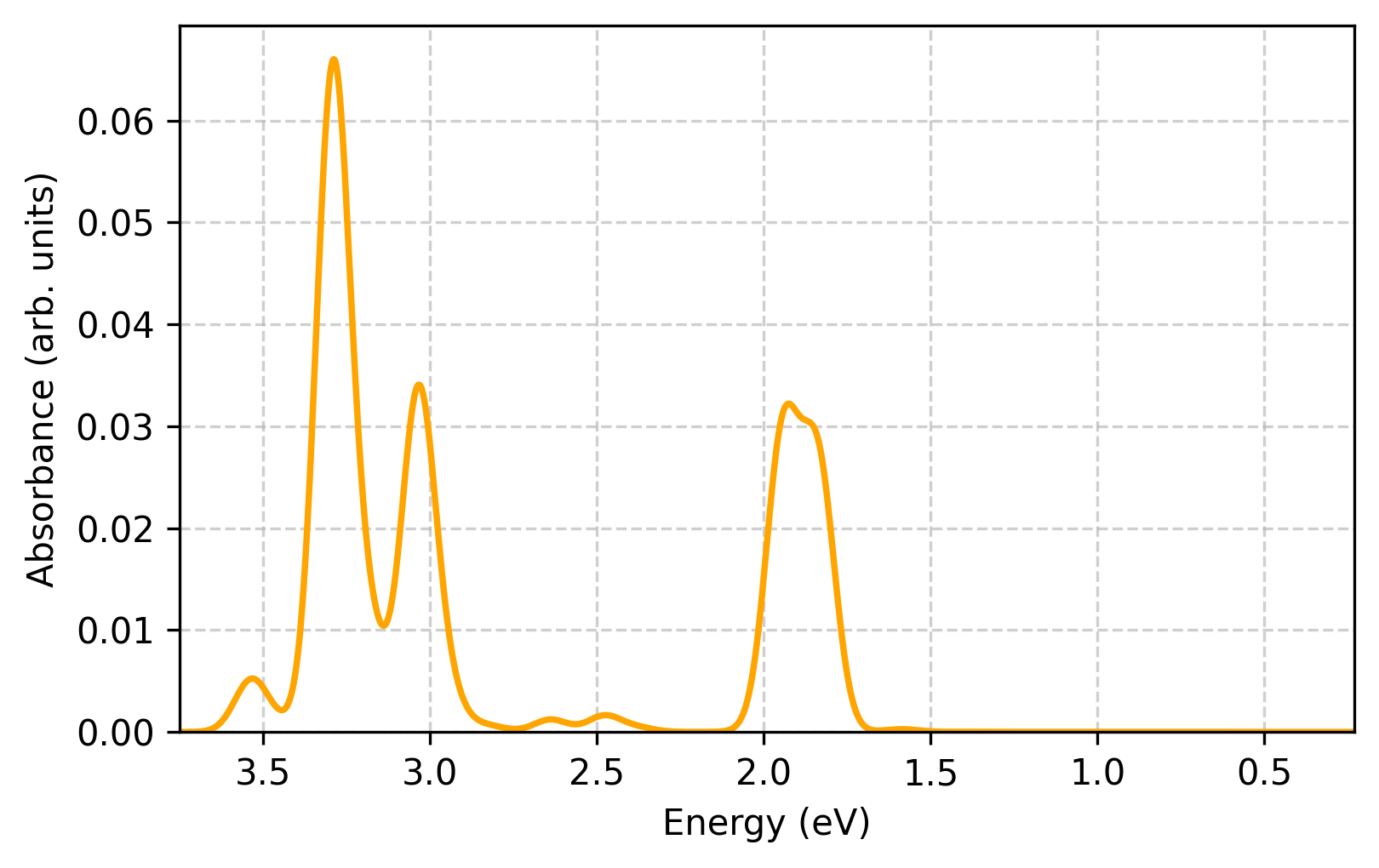}} \\
\end{tabular}
\caption{IR and UV-VIS spectra of B$_{18}$Ag$_2$ cluster.}
\label{fig_ir_vis}
\end{figure*}

The vibrational and electronic absorption properties of the B$_{18}$Ag$_2$ cluster were investigated to elucidate the interplay between structure, bonding, and spectroscopic response. As shown in Figure~\ref{fig_ir_vis}, the calculated infrared (IR) spectrum reflects this distinctive geometry and is characterized by a broad distribution of vibrational modes. In the low-frequency region below ~150~cm$^{-1}$, several modes arise from collective motions involving the Ag atoms oscillating relative to the two boron rings. A relatively intense band near 132~cm$^{-1}$ can be assigned to out-of-plane Ag–ring stretching, where the metal atoms move along the axis perpendicular to the boron layers, modulating the inter-ring interaction.

The mid-frequency region (400–800~cm$^{-1}$) dominates the IR spectrum and is strongly influenced by the double-ring topology. The most intense absorptions, located around 559, 588, and 633~cm$^{-1}$, correspond to collective skeletal vibrations of the two coupled boron rings. These modes involve synchronous and antisymmetric distortions of the rings, including in-plane breathing and inter-ring coupling motions. Additional intense features near 798 and 838~cm$^{-1}$ are associated with more localized but still cooperative deformations of the boron framework, highlighting the rigidity and delocalized bonding within the stacked rings. The presence of multiple closely spaced peaks in this region indicates symmetry lowering induced by Ag coordination, which lifts degeneracies that would otherwise be present in isolated boron rings.

At higher frequencies ($ > $ 900~cm$^{-1}$), the spectrum exhibits weaker and more localized modes corresponding primarily to B–B stretching vibrations within individual rings. Bands in the 1030–1070~cm$^{-1}$ and 1129–1210~cm$^{-1}$ regions can be attributed to intra-ring bond stretching, with limited participation of the Ag atoms. The relatively lower intensity of these modes reflects smaller dipole moment changes compared to the strongly coupled collective vibrations at lower frequencies.

The UV–Vis absorption spectrum of B$_{18}$Ag$_2$ spans a broad range from the near-infrared (NIR) to the near-ultraviolet (UV) region, reflecting a dense manifold of electronic excitations characteristic of its bent double-ring structure. The low-energy region below ~1.6 eV is dominated by very weak transitions with negligible oscillator strengths, indicating that these excitations are largely forbidden or involve limited transition dipole coupling. The onset of significant optical activity appears around ~1.8–2.0 eV, where moderately intense transitions emerge, followed by a pronounced absorption band in the visible region (~1.8–2.0 eV, 670–630 nm), with oscillator strengths reaching up to ~0.02–0.03. These features can be attributed to mixed metal–cluster excitations involving Ag-centered orbitals hybridized with the delocalized boron framework.

At higher energies, the spectrum becomes considerably more intense. Strong absorption bands are observed in the near-UV region between ~3.0 and 3.3 eV (~410–380 nm), where the oscillator strengths reach their maximum values (~0.03). These transitions are associated with highly allowed excitations, predominantly of $\pi \rightarrow \pi^*$ character, delocalized over the boron double-ring skeleton, with additional contributions from silver-induced electronic perturbations. The presence of multiple closely spaced intense transitions suggests partial degeneracy and a high density of states, consistent with the symmetry and extended conjugation of the bent double-ring geometry.

Overall, the optical response of B$_{18}$Ag$_2$ is governed by the interplay between the delocalized boron network and the embedded silver atoms, which act to modulate the electronic structure rather than simply donate charge. This results in a system that is weakly absorbing in the NIR but exhibits strong and structured absorption in the visible and near-UV regions, highlighting the role of Ag atoms in enhancing and redistributing the optical activity of the boron framework.


\subsection{Electron Density–Based Bonding Analysis}

\begin{table*}[ht!]
\centering
\begin{tabular}{cc}
 \resizebox*{0.40\textwidth}{!}{\includegraphics{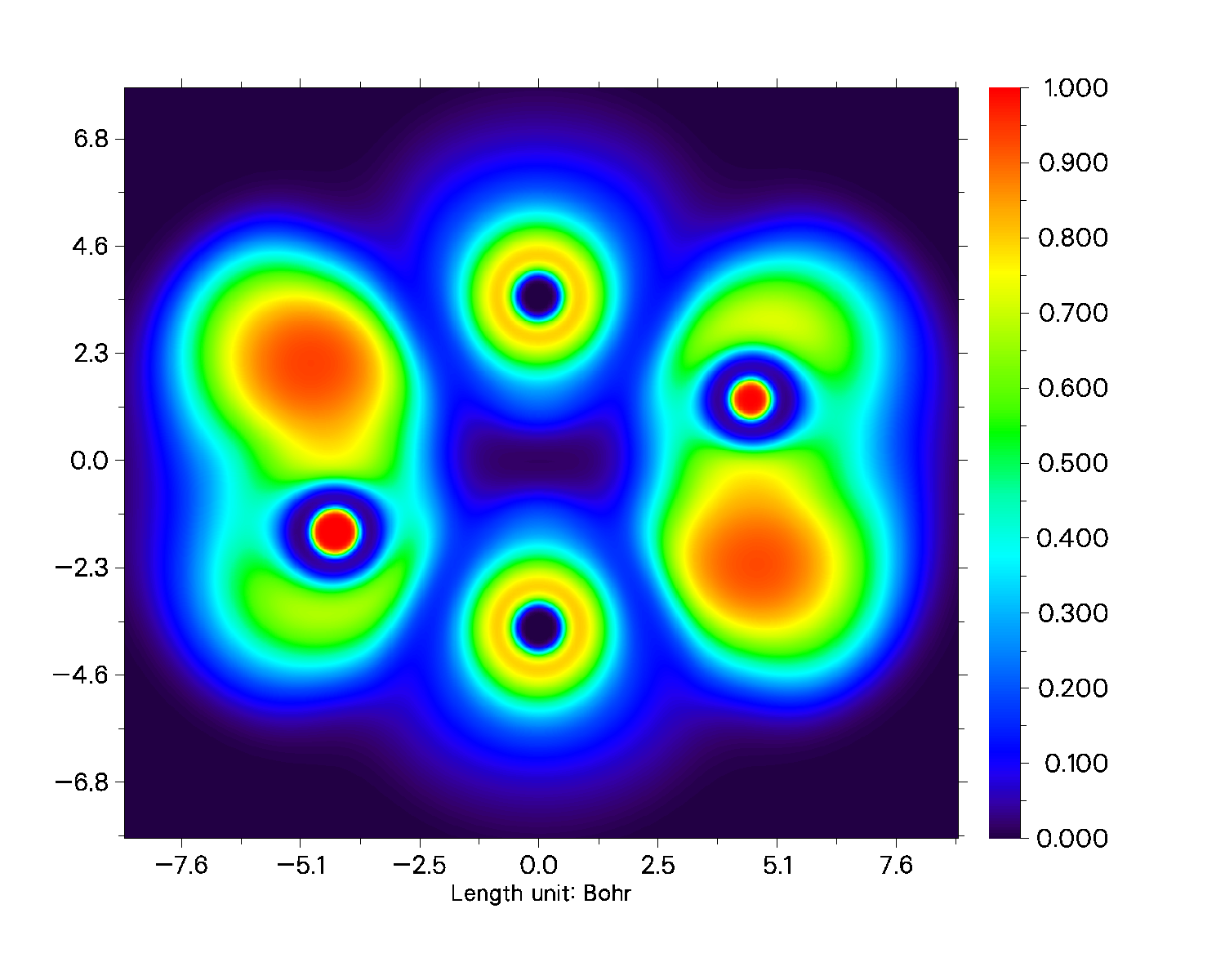}} &
\resizebox*{0.40\textwidth}{!}{\includegraphics{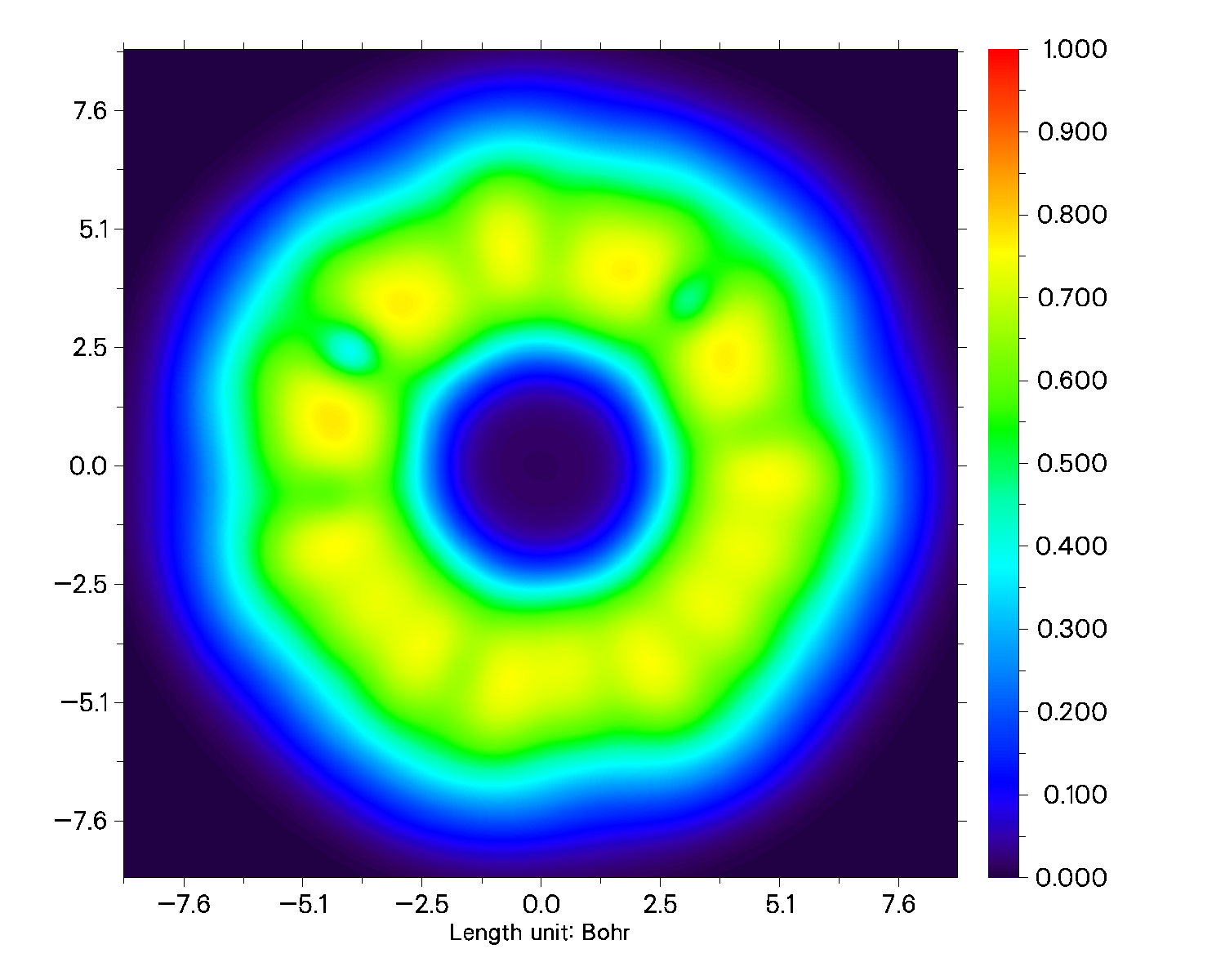}} \\
\end{tabular}
\caption{ELF 2D maps for the B$_{18}$Ag$_2$ cluster showing (left) the XZ-plane (side view) and (right) the XY-plane (top view), highlighting localized Ag regions and delocalized electron density over the boron framework.}
\label{fig_elf}
\end{table*}

The ELF maps of B$_{18}$Ag$_2$ reveal a bonding pattern characterized by the coexistence of localized Ag-associated basins and delocalized electron density over the boron framework (Figure~\ref{fig_elf}). In the XZ-plane (side view, left panel), the ELF distribution shows moderately localized regions around the Ag atoms, indicating weakly polarized Ag–B interactions. These features suggest limited charge transfer from Ag to the boron scaffold and a bonding scenario dominated by multicenter interactions rather than strongly directional Ag–B bonds. Additionally, the ELF between boron atoms appears partially connected, supporting the presence of multicenter B–B bonding across the stacked ring structure.

In contrast, the XY-plane (top view, right panel) displays a pronounced annular ELF distribution surrounding the B$_{18}$ double-ring framework. This nearly continuous ring-like pattern is characteristic of an extended $\sigma$-delocalized network, consistent with multicenter bonding and aromatic-like electron circulation along the boron skeleton. The reduced ELF density in the central cavity indicates the absence of inner-core bonding.

Overall, the bonding in B$_{18}$Ag$_2$ is governed by weak Ag–B interactions coupled with strong global $\sigma$-delocalization across the boron framework, which serves as the principal stabilizing factor for the tubular double-ring structure.

\begin{figure}[h!]
\centering
\resizebox*{0.48\textwidth}{!}{\includegraphics{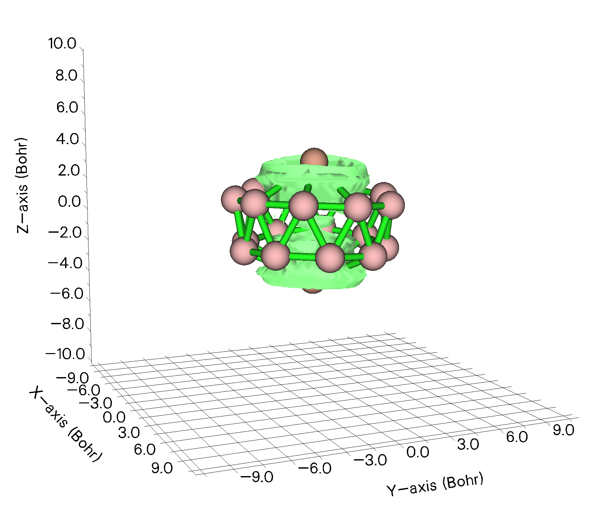}} \\
\caption{Reduced density gradient (RDG) isosurface for the B$_{18}$Ag$_2$ cluster (isovalue s=0.5).}
\label{fig_RDG}
\end{figure}

The reduced density gradient (RDG) isosurface of B$_{18}$Ag$_2$ further elucidates the nature of the interactions within the cluster (Figure~\ref{fig_RDG}). The isosurface is predominantly distributed over the boron framework, forming a continuous envelope around the double-ring structure, which is indicative of extensive multicenter interactions among the boron atoms. The absence of highly localized, sharply defined RDG features between specific atom pairs suggests that the bonding within the boron skeleton is largely delocalized rather than dominated by conventional two-center bonds. In the regions associated with the Ag atoms, the RDG isosurfaces appear more diffuse and less pronounced, consistent with weak, noncovalent or partially attractive Ag–B interactions. This behavior supports the notion that silver does not form strong directional bonds with the boron framework but instead interacts through weakly attractive forces and electrostatic contributions. Overall, the RDG analysis corroborates the ELF results, confirming that the stability of B$_{18}$Ag$_2$ arises primarily from delocalized multicenter bonding within the boron network, complemented by weak Ag–B interactions.

\begin{figure}[h!]
\centering
\resizebox*{0.25\textwidth}{!}{\includegraphics{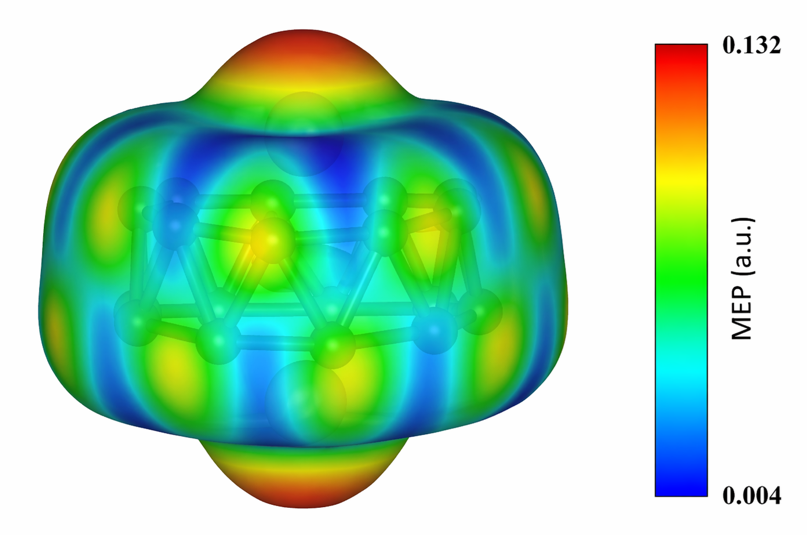}} \\
\caption{Molecular electrostatic potential (MEP) mapped onto the electron density isosurface ($\rho$ = 0.01 a.u.) for the B$_{18}$Ag$_2$ cluster.}
\label{mep}
\end{figure}

The molecular electrostatic potential (MEP) map of B$_{18}$Ag$_2$ provides further insight into the charge distribution and electrostatic characteristics of the cluster (Figure~\ref{mep}).\cite{doi:10.1021/acs.jpca.4c08443,D0CP04018E,D6CP00681G} The surface exhibits regions of negative potential predominantly localized over the boron framework, indicating electron-rich domains associated with the delocalized bonding network of the double-ring structure. In contrast, the regions around the Ag atoms display relatively positive electrostatic potential, reflecting electron deficiency and suggesting partial charge transfer from silver to the boron scaffold. This polarization is consistent with the weakly interacting nature of the Ag–B contacts inferred from the RDG analysis. The overall MEP distribution highlights a clear separation between electron-rich boron regions and electron-poor metallic sites, supporting a bonding picture in which the boron framework acts as an electron reservoir stabilized by global delocalization, while the Ag atoms contribute primarily through electrostatic interactions rather than strong covalent bonding.

\section{Conclusions}

In this work, the structural stability, electronic properties, and bonding characteristics of the B$_{18}$Ag$_2$ cluster were systematically investigated using density functional theory combined with global optimization techniques. Basin-hopping searches identify a bent double-ring structure as the global minimum, consisting of two stacked B$_9$ rings stabilized by Ag atoms positioned above and below the boron framework. Vibrational analyses confirm the dynamical stability of the low-energy structures, while the calculated IR and UV–Vis spectra reveal characteristic collective vibrational modes and delocalized electronic excitations associated with the boron framework and modulated by Ag incorporation. Real-space bonding analyses based on ELF, RDG, and MEP consistently demonstrate that the stability of B$_{18}$Ag$_2$ is governed by extensive multicenter $\sigma$-delocalization over the boron skeleton. In contrast, Ag–B interactions are weak, largely non-directional, and primarily electrostatic, playing a secondary role as structural stabilizers and electronic modulators. The continuous annular electron delocalization within the double-ring structure is consistent with an aromatic-like bonding pattern. Overall, B$_{18}$Ag$_2$ can be described as a silver-stabilized boron double-ring cluster in which global electron delocalization dictates structural stability, while Ag atoms act as axial centers that fine-tune the electronic structure. These findings provide new insight into the role of coinage-metal doping in stabilizing extended boron nanostructures and may guide the design of novel boron-based materials with tunable electronic properties.


\section{Acknowledgments}
P.L.R.-K. would like to thank the support of CIMAT Supercomputing Laboratories of Guanajuato and Puerto Interior. 



\bibliographystyle{unsrt}
\bibliography{mendelei.bib}

@article{doi:10.1021/ct100199k,
author = {Kossmann, Simone and Neese, Frank},
title = {Efficient Structure Optimization with Second-Order Many-Body Perturbation Theory: The RIJCOSX-MP2 Method},
journal = {Journal of Chemical Theory and Computation},
volume = {6},
number = {8},
pages = {2325-2338},
year = {2010},
doi = {10.1021/ct100199k},
note ={PMID: 26613489},
URL = {https://doi.org/10.1021/ct100199k},
eprint = {https://doi.org/10.1021/ct100199k}
}

@Article{D6CP00681G,
author ="Guevara-Vela, José Manuel and Rodríguez-Kessler, Peter L. and Muñoz-Castro, Alvaro and Rocha-Rinza, Tomás",
title  ="{Global minimum structures and electronic stability of Pt-doped silicon clusters PtSin (n = 2 to 11) in neutral and anionic charge states}",
journal  ="Phys. Chem. Chem. Phys.",
year  ="2026",
pages  ="-",
publisher  ="The Royal Society of Chemistry",
doi  ="10.1039/D6CP00681G",
url  ="http://dx.doi.org/10.1039/D6CP00681G"}

@article{doi:10.1126/science.196.4294.1047,
author = {William N. Lipscomb },
title = {The Boranes and Their Relatives},
journal = {Science},
volume = {196},
number = {4294},
pages = {1047-1055},
year = {1977},
doi = {10.1126/science.196.4294.1047},
URL = {https://www.science.org/doi/abs/10.1126/science.196.4294.1047},
eprint = {https://www.science.org/doi/pdf/10.1126/science.196.4294.1047}}

@Article{D0CP04018E,
author ="Rodríguez-Kessler, Peter L. and Rodríguez-Domínguez, Adán R. and MacLeod Carey, Desmond and Muñoz-Castro, Alvaro",
title  ="Structural characterization{,} reactivity{,} and vibrational properties of silver clusters: a new global minimum for Ag$_{16}$",
journal  ="Phys. Chem. Chem. Phys.",
year  ="2020",
volume  ="22",
issue  ="46",
pages  ="27255-27262",
publisher  ="The Royal Society of Chemistry",
doi  ="10.1039/D0CP04018E",
url  ="http://dx.doi.org/10.1039/D0CP04018E"}

@article{doi:10.1021/acs.jpca.4c08443,
author = {Guevara-Vela, J. M. and Rodríguez-Kessler, P. L. and Cabellos-Quiroz, J. L. and Rocha-Rinza, T. and Vásquez-Espinal, A. and Mu{\~n}oz-Castro, A.},
title = "{Structure- and Size-Dependent Properties of B$_n$Cu$_2^{0/-}$ (n = 2-14) Clusters: DFT Calculations}",
journal = {The Journal of Physical Chemistry A},
volume = {129},
number = {22},
pages = {4855-4860},
year = {2025},
doi = {10.1021/acs.jpca.4c08443},
URL = {    https://doi.org/10.1021/acs.jpca.4c08443},
eprint = {    https://doi.org/10.1021/acs.jpca.4c08443}
}

@Article{Ortega-Flores2025,
author={Ortega-Flores, Samantha
and Rodr{\'i}guez-Kessler, P. L.
and Mu{\~{n}}oz-Castro, Alvaro},
title="{Size-dependent structural motifs in Ag$_n$Mo (n = 1-14) clusters: from planar to icosahedral architectures}",
journal={Journal of Nanoparticle Research},
year={2025},
month={Oct},
day={13},
volume={27},
number={10},
pages={277},
issn={1572-896X},
doi={10.1007/s11051-025-06471-3},
url={https://doi.org/10.1007/s11051-025-06471-3}
}

@Article{Rodriguez-Kessler2026,
author={Rodr{\'i}guez-Kessler, P. L. and Rodr{\'i}guez-Dom{\'i}nguez, A. R. and Mu{\~{n}}oz-Castro, Alvaro},
title="{Size-dependent structural motifs in Pt$_n$V$^-$(n = 1--13) cluster anions: a DFT insight}",
journal={Journal of Nanoparticle Research},
year={2026},
month={Feb},
day={20},
volume={28},
number={3},
pages={59},
issn={1572-896X},
doi={10.1007/s11051-026-06580-7},
url={https://doi.org/10.1007/s11051-026-06580-7}
}

@Article{D5SC08598E,
author ="Choi, Hyun Wook and Kahraman, Deniz and Chen, Wei-Jia and Wang, Lai-Sheng",
title  ="Probing the weak interaction between silver and boron",
journal  ="Chem. Sci.",
year  ="2026",
volume  ="17",
issue  ="10",
pages  ="5084-5091",
publisher  ="The Royal Society of Chemistry",
doi  ="10.1039/D5SC08598E",
url  ="http://dx.doi.org/10.1039/D5SC08598E",
}

@article{GUEVARAVELA2025115487,
title = "{Structure search for B$_7$Mn$_2$ clusters: Inverse sandwich geometry with a high-spin state}",
journal = {Computational and Theoretical Chemistry},
volume = {1254},
pages = {115487},
year = {2025},
issn = {2210-271X},
doi = {https://doi.org/10.1016/j.comptc.2025.115487},
url = {https://www.sciencedirect.com/science/article/pii/S2210271X25004232},
author = {Jose Manuel Guevara-Vela and J.L. Cabellos-Quiroz and Sebastián Salazar-Colores and P.L. Rodríguez-Kessler and Alvaro Muñoz-Castro},
}

@Article{D5CP01078K,
author ="Rodríguez-Kessler, Peter L. and Muñoz-Castro, Alvaro",
title  ="{Al$_2$B$_7$: an extension of the inverse sandwich B$_9$ cluster featuring Lewis acid sites and planar aromaticity}",
journal  ="Phys. Chem. Chem. Phys.",
year  ="2025",
volume  ="27",
issue  ="24",
pages  ="12793-12800",
publisher  ="The Royal Society of Chemistry",
doi  ="10.1039/D5CP01078K",
url  ="http://dx.doi.org/10.1039/D5CP01078K"}

@Article{C7CP04158F,
author ="Muñoz-Castro, Alvaro and Popov, Ivan A. and Boldyrev, Alexander I.",
title  ="{Long-range magnetic response of toroidal boron structures: B16 and [Co@B16]-/3- species}",
journal  ="Phys. Chem. Chem. Phys.",
year  ="2017",
volume  ="19",
issue  ="38",
pages  ="26145-26150",
publisher  ="The Royal Society of Chemistry",
doi  ="10.1039/C7CP04158F",
url  ="http://dx.doi.org/10.1039/C7CP04158F"}

@Article{C6SC02623K,
author ="Jian, Tian and Li, Wan-Lu and Chen, Xin and Chen, Teng-Teng and Lopez, Gary V. and Li, Jun and Wang, Lai-Sheng",
title  ="{Competition between drum and quasi-planar structures in RhB$_{18}^-$: motifs for metallo-boronanotubes and metallo-borophenes}",
journal  ="Chem. Sci.",
year  ="2016",
volume  ="7",
issue  ="12",
pages  ="7020-7027",
publisher  ="The Royal Society of Chemistry",
doi  ="10.1039/C6SC02623K",
url  ="http://dx.doi.org/10.1039/C6SC02623K"}

@article{doi:10.1021/jp8087918,
author = {Johansson, Mikael P.},
title = "{On the Strong Ring Currents in B20 and Neighboring Boron Toroids}",
journal = {The Journal of Physical Chemistry C},
volume = {113},
number = {2},
pages = {524-530},
year = {2009},
doi = {10.1021/jp8087918},
URL = {  https://doi.org/10.1021/jp8087918},
eprint = {  https://doi.org/10.1021/jp8087918
}
}

@article{doi:10.1021/acs.jpclett.0c02656,
author = {Lu, Cheng and Gong, Weiguang and Li, Quan and Chen, Changfeng},
title = "{Elucidating Stress–Strain Relations of ZrB12 from First-Principles Studies}",
journal = {The Journal of Physical Chemistry Letters},
volume = {11},
number = {21},
pages = {9165-9170},
year = {2020},
doi = {10.1021/acs.jpclett.0c02656},
   URL = { https://doi.org/10.1021/acs.jpclett.0c02656},
eprint = {   https://doi.org/10.1021/acs.jpclett.0c02656
}
}

@article{doi:10.1021/acs.inorgchem.7b02585,
author = {Chen, Bo Le and Sun, Wei Guo and Kuang, Xiao Yu and Lu, Cheng and Xia, Xin Xin and Shi, Hong Xiao and Maroulis, George},
title = "{Structural Stability and Evolution of Medium-Sized Tantalum-Doped Boron Clusters: A Half-Sandwich-Structured TaB$_{12}^-$ Cluster}",
journal = {Inorganic Chemistry},
volume = {57},
number = {1},
pages = {343-350},
year = {2018},
doi = {10.1021/acs.inorgchem.7b02585},
   URL = {  https://doi.org/10.1021/acs.inorgchem.7b02585},
eprint = {  https://doi.org/10.1021/acs.inorgchem.7b02585}
}

@article{RODRIGUEZKESSLER2023116538,
title = "{Structure and stability of Cu-doped Bn (n = 1-12) clusters: DFT calculations}",
journal = {Polyhedron},
volume = {243},
pages = {116538},
year = {2023},
issn = {0277-5387},
doi = {https://doi.org/10.1016/j.poly.2023.116538},
url = {https://www.sciencedirect.com/science/article/pii/S0277538723002607},
author = {P.L. Rodríguez-Kessler and Alejandro Vásquez-Espinal and Alvaro Muñoz-Castro}
}

@article{RODRIGUEZKESSLER2024122062,
title = "{Structures of Ni-doped Bn (n = 1-13) clusters: A computational study}",
journal = {Inorganica Chimica Acta},
volume = {568},
pages = {122062},
year = {2024},
issn = {0020-1693},
doi = {https://doi.org/10.1016/j.ica.2024.122062},
url = {https://www.sciencedirect.com/science/article/pii/S002016932400152X},
author = {P.L. Rodríguez-Kessler and Alejandro Vásquez-Espinal and A.R. Rodríguez-Domínguez and J.L. Cabellos-Quiroz and A. Muñoz-Castro}
}

@article{RODRIGUEZKESSLER2025117486,
title = "{Structure search for B$_7$Cr$_2$ clusters: Inverse sandwich structure for the global minimum}",
journal = {Polyhedron},
volume = {273},
pages = {117486},
year = {2025},
issn = {0277-5387},
doi = {https://doi.org/10.1016/j.poly.2025.117486},
url = {https://www.sciencedirect.com/science/article/pii/S0277538725001007},
author = {P.L. Rodríguez-Kessler and Alvaro Muñoz-Castro},
}

@Article{Zhuan-Yu2014,
author={Zhuan-Yu, Wang
and Wei-Li, Kang
and Jian-Feng, Jia
and Hai-Shun, Wu},
title="{Structure and stability of Ti$_2$B$_n$ (n=1-10) clusters: an ab initio investigation}",
journal={Acta Physica Sinica},
year={2014},
volume={63},
number={23},
pages={233102-233102},
doi={10.7498/aps.63.233102},
url={https://wulixb.iphy.ac.cn/en/article/doi/10.7498/aps.63.233102},
url={https://doi.org/10.7498/aps.63.233102}
}

@article{JIA2014128,
title = "{Density functional theory investigation on the structure and stability of Sc$_2$B$_n$ (n=1–10) clusters}",
journal = {Computational and Theoretical Chemistry},
volume = {1027},
pages = {128-134},
year = {2014},
issn = {2210-271X},
doi = {https://doi.org/10.1016/j.comptc.2013.11.008},
url = {https://www.sciencedirect.com/science/article/pii/S2210271X13004921},
author = {Jianfeng Jia and Xiaorong Li and Yanan Li and Lijuan Ma and Hai-Shun Wu}
}

@article{OLALDELOPEZ2024,
title = "{Hydrogen storage properties for bimetallic doped boron clusters M$_2$B$_7$ (M=Fe, Co, Ni)}",
journal = {International Journal of Hydrogen Energy},
year = {2024},
issn = {0360-3199},
doi = {https://doi.org/10.1016/j.ijhydene.2024.05.429},
url = {https://www.sciencedirect.com/science/article/pii/S0360319924021396},
author = {David Olalde-López and P.L. Rodríguez-Kessler and Salomón Rodríguez-Carrera and A. Muñoz-Castro},
}

@article{LI202325821,
title = "{Geometric structures and hydrogen storage properties of M$_2$B$_7$ (M=Be, Mg, Ca) clusters}",
journal = {International Journal of Hydrogen Energy},
volume = {48},
number = {66},
pages = {25821-25829},
year = {2023},
issn = {0360-3199},
doi = {https://doi.org/10.1016/j.ijhydene.2023.03.213},
!url = {https://www.sciencedirect.com/science/article/pii/S0360319923012831},
author = {Hai-Ru Li and Ceng Zhang and Wan-Biao Ren and Ying-Jin Wang and Tao Han},
keywords = {Density functional theory, Hydrogen storage density, Reversible adsorption, Hydrogen storage material},
abstract = {Based on the density functional theory, we investigate the electronic properties of the clusters M2B7 (M = Be, Mg, Ca) and their hydrogen storage properties systematically in this paper. Extensive global search results show that the global minimal structures of the three systems (Be2B7, Mg2B7 and Ca2B7) are heptagonal biconical structure, and the two alkaline earth metals are located at the top of the biconical. Chemical bonding analyses show that M2B7 clusters have 6σ and 6π delocalized electrons, which are doubly aromatic. At the wB97XD level, the three systems have good hydrogen storage capabilities. The hydrogen storage density of Be2B7 is as high as 23.03 wt%, while Mg2B7 and Ca2B7 also far exceed the hydrogen storage target set by the U.S. Department of Energy in 2017. Their average adsorption energies of H2 molecules all ranged from 0.1 eV/H2 to 0.48 eV/H2, which is fall in between physisorption and chemisorption. Extensive Born Oppenheimer molecular dynamics (BOMD) simulations show that the H2 molecules of the three systems can be completely released at a certain temperature. Therefore, M2B7 systems can achieve reversible adsorption of H2 molecules at normal temperature and pressure. It can be seen that the B7 clusters modified by alkaline earth metals may become a promising new nano-hydrogen storage material.}
}

@Article{C5CP01650A,
author ="Pham, Hung Tan and Nguyen, Minh Tho",
title  ="{Effects of bimetallic doping on small cyclic and tubular boron clusters: B$_7$M$_2$ and B$_{14}$M$_2$ structures with M = Fe{,} Co}",
journal  ="Phys. Chem. Chem. Phys.",
year  ="2015",
volume  ="17",
issue  ="26",
pages  ="17335-17345",
publisher  ="The Royal Society of Chemistry",
doi  ="10.1039/C5CP01650A",
url  ="http://dx.doi.org/10.1039/C5CP01650A",
}

@article{PHAM2019186,
title = "{Impressive capacity of the B$_7^-$ and V$_2$B$_7$ clusters for CO$_2$ capture}",
journal = {Chemical Physics Letters},
volume = {728},
pages = {186-194},
year = {2019},
issn = {0009-2614},
doi = {https://doi.org/10.1016/j.cplett.2019.04.087},
url = {https://www.sciencedirect.com/science/article/pii/S0009261419303823},
author = {Hung Tan Pham and My Phuong Pham-Ho and Minh Tho Nguyen}
}

@article{10.1063/5.0004608,
    author = {Neese, Frank and Wennmohs, Frank and Becker, Ute and Riplinger, Christoph},
    title = "{The ORCA quantum chemistry program package}",
    journal = {The Journal of Chemical Physics},
    volume = {152},
    number = {22},
    pages = {224108},
    year = {2020},
    month = {06},
       issn = {0021-9606},
    doi = {10.1063/5.0004608},
    url = {https://doi.org/10.1063/5.0004608},
    eprint = {https://pubs.aip.org/aip/jcp/article-pdf/doi/10.1063/5.0004608/16740678/224108\_1\_online.pdf},
}

@article{10.1063/1.478522,
    author = {Adamo, Carlo and Barone, Vincenzo},
    title = "{Toward reliable density functional methods without adjustable parameters: The PBE0 model}",
    journal = {The Journal of Chemical Physics},
    volume = {110},
    number = {13},
    pages = {6158-6170},
    year = {1999},
    month = {04},
    issn = {0021-9606},
    doi = {10.1063/1.478522},
    url = {https://doi.org/10.1063/1.478522},
    eprint = {https://pubs.aip.org/aip/jcp/article-pdf/110/13/6158/10797469/6158\_1\_online.pdf},
}

@Article{B508541A,
author ="Weigend, Florian and Ahlrichs, Reinhart",
title  ="{Balanced basis sets of split valence{,} triple zeta valence and quadruple zeta valence quality for H to Rn: Design and assessment of accuracy}",
journal  ="Phys. Chem. Chem. Phys.",
year  ="2005",
volume  ="7",
issue  ="18",
pages  ="3297-3305",
publisher  ="The Royal Society of Chemistry",
doi  ="10.1039/B508541A",
url  ="http://dx.doi.org/10.1039/B508541A"}

@article{https://doi.org/10.1002/jcc.22885,
author = {Lu, Tian and Chen, Feiwu},
title = {Multiwfn: A multifunctional wavefunction analyzer},
journal = {Journal of Computational Chemistry},
volume = {33},
number = {5},
pages = {580-592},
doi = {https://doi.org/10.1002/jcc.22885},
url = {https://onlinelibrary.wiley.com/doi/abs/10.1002/jcc.22885},
eprint = {https://onlinelibrary.wiley.com/doi/pdf/10.1002/jcc.22885},
year = {2012}
}

\end{document}